\documentclass[%
 preprint,
 amsmath,amssymb,amsthm,
 prl,
 longbibliography,
superscriptaddress,
]{revtex4-2}

\usepackage{graphicx}
\usepackage{dcolumn}
\usepackage{bm}
\usepackage{color}
\usepackage{float}

\begin{document}

\title{Controlling quasi-parametric amplifications: From multiple $\mathcal{PT}$-symmetry phase transitions to non-Hermitian sensing}

\author{Xiaoxiong Wu}
\affiliation{State Key Laboratory of Advanced Optical Communication Systems and Networks, School of Physics and Astronomy, Shanghai Jiao Tong University, Shanghai 200240, China}

\author{Kai Bai}
\affiliation{Key Laboratory of Artificial Micro- and Nano-structures of Ministry of Education and School of Physics and Technology, Wuhan University, Wuhan 430072, China}

\author{Penghong Yu}
\affiliation{State Key Laboratory of Advanced Optical Communication Systems and Networks, School of Physics and Astronomy, Shanghai Jiao Tong University, Shanghai 200240, China}

\author{Zhaohui Dong}
\affiliation{State Key Laboratory of Advanced Optical Communication Systems and Networks, School of Physics and Astronomy, Shanghai Jiao Tong University, Shanghai 200240, China}

\author{Yanyan He}
\affiliation{State Key Laboratory of Advanced Optical Communication Systems and Networks, School of Physics and Astronomy, Shanghai Jiao Tong University, Shanghai 200240, China}

\author{Jingui Ma}
\affiliation{Key Laboratory for Laser Plasmas (MOE), Collaborative Innovation Center of IFSA (CICIFSA), School of Physics and Astronomy, Shanghai Jiao Tong University, Shanghai 200240, China}

\author{Vladislav V. Yakovlev}
\affiliation{Texas A\&M University, College Station, TX 77843, USA}

\author{Meng Xiao}
\email{phmxiao@whu.edu.cn}
\affiliation{Key Laboratory of Artificial Micro- and Nano-structures of Ministry of Education and School of Physics and Technology, Wuhan University, Wuhan 430072, China}
\affiliation{Wuhan Institute of Quantum Technology, Wuhan 430206, China}

\author{Xianfeng Chen}
\affiliation{State Key Laboratory of Advanced Optical Communication Systems and Networks, School of Physics and Astronomy, Shanghai Jiao Tong University, Shanghai 200240, China}
\affiliation{Jinan Institute of Quantum Technology, Jinan 250101, China}
\affiliation{Collaborative Innovation Center of Light Manipulation and Applications, Shandong Normal University, Jinan 250358, China}

\author{Luqi Yuan}
\email{yuanluqi@sjtu.edu.cn}
\affiliation{State Key Laboratory of Advanced Optical Communication Systems and Networks, School of Physics and Astronomy, Shanghai Jiao Tong University, Shanghai 200240, China}

\date{\today}

\begin{abstract}

Quasi-parametric amplification (QPA) is a nonlinear interaction in which the idler wave is depleted through some loss mechanism. QPA plays an important role in signal amplification in ultrafast photonics and quantum light generation. The QPA process has a number of features characterized by the non-Hermitian parity-time ($\mathcal{PT}$) symmetry. In this report, we explore new interaction regimes and uncover multiple $\mathcal{PT}$-symmetry phase transitions in such QPA process where transitions are particularly sensitive to external parameters. In particular, we demonstrate the feasibility of detection of $10^{-11}$ inhomogeneities of the doped absorber, which is order of magnitude more sensitive than similar measurements performed in a linear absorption regime. In doing so, we reveal a family of $\mathcal{PT}$-symmetry phase transitions appearing in the QPA process and provide a novel nonlinear optical sensing mechanism for precise optical measurements.

\end{abstract}

\maketitle
\newpage


\section{Introduction}
In the past decade, non-Hermiticity introduced by the nonreciprocal coupling or exchange with surrounding environment has ignited tremendous research interest, and explorations of non-Hermitian physics using light or waves have attracted flourishing attention where various exotic phenomena and many unique features different from those in conventional Hermitian systems are explored \cite{feng2017npho,zhao2018nsr,el2018nphy,ozdemir2019nm,assawaworrarit2020ne,
parto2021nanop,ding2022nrp,yan2023nanop,cheng2022PRB,wu2023APR}.  One of the key findings is the exceptional point (EP) surpporting parity-time ($\mathcal{PT}$) symmetry with balanced loss and gain distributions in space, i.e., $[\mathcal{PT},H_{PT}]=0$, where multiple bands coalesce with each other at EP, and afterwards fruitful applications have been proposed including unidirectional invisibility \cite{lin2011prl,yin2013naturem,huang2017nanoph,zhou2023prl} and single-mode lasing \cite{zhu2022prl,schumer2022science,li2023nature}. Due to the extreme responsivity of the EP from the surrounding change of environment \cite{hodaei2017nature,chen2017nature,miri2019science,Wiersig2020pr,peters2022prl}, it has been proposed that one can use EP to perform sensing in photonics \cite{feng2023prl,xu2023sciadv}. Although there are some debates on the efficiency \cite{hodaei2017nature,langbein2018PRA,ozdemir2019nm,chen2019NJP,bai2022nsr,bai2024prl}, the idea of using its nonlinear dependence of parameters near EP is fundamentally intriguing. 

Most of previous studies on the non-Hermitian EP sensing fall in photonic platforms. Recently it has been noticed that the nonlinear wave-mixing process in the ultrafast science may also support the non-Hermitian engineering, where the property of EP has been discussed \cite{flemens2021oe,flemens2022prl,chen2023Ultrafast,ni2023AP}. In particular, the connection between non-Hermiticity and the nonlinear wave-mixing processes such as optical parametric amplifications has been unveiled \cite{el2015ol,zhong2016njp,miri2016njp,el2019cp,flemens2021oe,chen2021Elight,flemens2022prl},
which provides deeper understanding on the nonlinear photon conversion and proposes the quasi-parametric amplification (QPA) with high nonlinear conversion efficiency  \cite{ma2015optica,el2015ol,zhong2016njp,miri2016njp,ma2017oe,el2019cp}. Yet, so far, the behaviors of phase transitions and EP in this QPA system still lack further elaborate studies, especially for the dynamical evolution and its potential application with the use of the $\mathcal{PT}$-symmetry phase transition such as aforementioned sensing. Especially, high-precision sensing has remarkable boost for scientific researches \cite{andrew1986RMP,degen2017RMP}, while QPA system may serve a unique candidate to detect the sensing information with very weak fluctuations.

In this report, we investigate a broad class of novel QPA systems including the parametric down-conversion (PDC) process that generates signal photons and linear depletion that absorbs idler photons. This hybrid process can be associated with the non-Hermitian $\mathcal{PT}$-symmetry system dependent on the signal photon flux density along the propagation distance. Under different parametric conditions, multiple $\mathcal{PT}$-symmetry phase transitions exist, which results from the competition between the absorption of the idler photon and the signal amplification from PDC. We find that multiple transitions around EPs provides a dramatic variation of signal photons with sharp dips. Such unique feature can be utilized for a sensing detection to measure the homogeneousness of the doped absorber, i.e., very small perturbations can be measured with ultra-sensitivity, which can be verified in our simulations. Thus, our work opens a new perspective on the relationship between the nonlinear wave-mixing process and non-Hermiticity and unpacked an important avenue for the ultra-sensitive sensing at EPs in nonlinear optics.

\section{Model and theory}

\begin{figure}[H]
\center
\includegraphics[width=12cm]{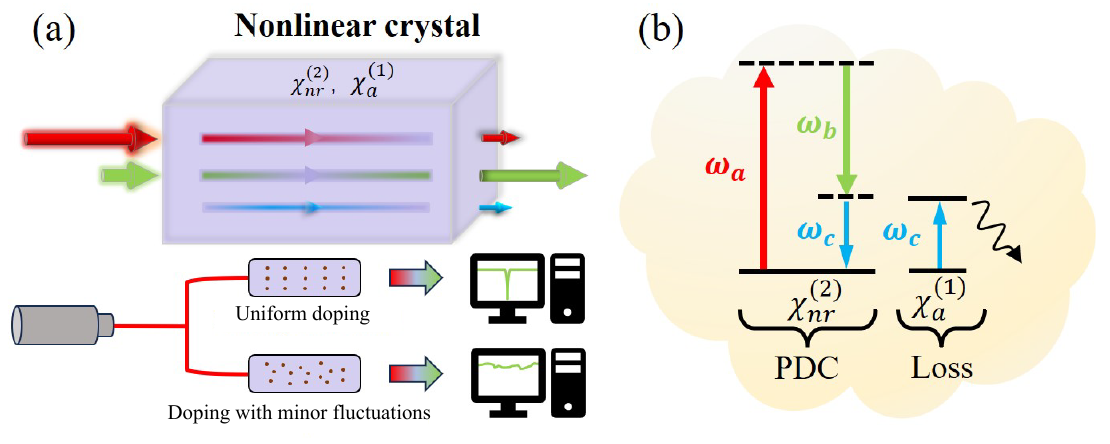}
\caption{\label{fig1} Principle of QPA based on nonlinear PDC and linear absorption. (a) Schematic of a typical PDC process where the red, green and blue arrows represent pump ($\omega_a$), signal ($\omega_b$) and idler beams ($\omega_c$), respectively. The pump, signal, and/or idler beams are injected into the nonlinear crystal which includes both the second-order nonlinearity $\chi^{(2)}_{nr}$ and the linear absorber with $\chi_\alpha^{(1)}$. The panel below the schematic represents measuring intensities for crystal either with the uniform doping or with the doping with minor fluctuations, which reveals the sensing concept using EPs. (b) Illustration of PDC including a conversion between one pump photon $\omega_a$, one signal photon $\omega_b$ and one idler photon $\omega_c$, as well as the linear absorption of the idler photon.}
\end{figure} 


We consider a PDC process where light is injected into a nonlinear optical crystal with the second-order nonlinearity  $\chi^{(2)}_{nr}$, doped with linear absorbers with $\chi_a^{(1)}$. As depicted in Fig.~\ref{fig1}, such second-order nonlinear process can support the photon conversion from a pump photon at the frequency $\omega_a$ to two photons, i.e., the signal photon at $\omega_b$ and the idler photon at $\omega_c$ ($\omega_a=\omega_b+\omega_c$), and vice versa. The doped absorber is chosen to have the energy transition close to $\omega_c$, so it plays the role of loss for photons at $\omega_c$. Such linear depletion of idler photons can lead to QPA if one injects strong pump beam together with a weak signal probe, where the nonlinear process can be operated only inside the $\mathcal{PT}$-broken phase to reach the high-efficient generation of the signal \cite{ma2015optica,el2015ol,miri2016njp,zhong2016njp,el2019cp,flemens2022prl}. In this work, different from Ref.~\cite{ma2015optica,el2015ol,miri2016njp,zhong2016njp,el2019cp,flemens2022prl}, we stick ourselves to this system but explore the $\mathcal{PT}$-symmetry phase transitions based on different combination of inject beams and then point out the potential for sensing in nonlinear optical processes.

For the light propagation along the $z$ direction in the nonlinear crystal, the electric field can be described by \cite{miri2016njp}
\begin{equation}\label{eq1}
\begin{aligned}\nabla^2 E(x,z,t)-\frac{{n_r}^2}{c_0^2}\frac{\partial^2}{\partial t^2}E(x,z,t)=\frac{1}{\epsilon_0 c_0^2}\frac{\partial^2P_{NL}}{\partial t^2},\end{aligned}
\end{equation}
where $c_0$ is the speed of light in vacuum, $n_r$ describes the refractive index, and $P_{NL}=\epsilon_0 d_{eff}|E|^2$ is the nonlinear polarization with the second-order nonlinear coefficient $d_{eff}$ and dielectric constant in vacuum $\epsilon_0$. The electric field can be further expanded as $E=\frac{1}{2}\sum\limits_j A_j(z)B_j(x)e^{i(\omega_j t-\beta_j z)} + \rm{c.c.}$ where $j=a,b,c$, $\beta_j=n_r(\omega_j)\beta_0$ giving the wave vector at the frequency $\omega_j$, $B_j(x)$ is the transverse modal profile, and $A_j(z)$ is the field amplitude for the pump, signal, or idler field respectively \cite{zhong2016njp,miri2016njp}. Then, together with the slowly varying envelope approximation and the rotating-wave approximation, Eq.~(\ref{eq1}) becomes \cite{el2015ol,ma2017oe,flemens2022prl} 
\begin{equation}\label{eq2}
\left\{ \begin{array}{l}
{d u_a(z)}/{dz} = i{\Gamma}{u_b}(z){u_c}(z)e^{-i\Delta\beta z},\\
{d u_b(z)}/{dz} = i{\Gamma}{u_a}(z)u_c^*(z)e^{i\Delta\beta z},\\
{d u_c(z)}/{dz}= i{\Gamma}{u_a}(z)u_b^*(z)e^{i\Delta\beta z} - \alpha {u_c}(z),
\end{array} \right.
\end{equation}
where $\Gamma$ represents the nonlinear coupling coefficient of the PDC process, $\alpha\ge 0$ is the linear absorption coefficient, $\Delta\beta=\beta_a-(\beta_b+\beta_c)$, and $u_j(z)\equiv \sqrt{\frac{2 n_r(\omega_j)\epsilon_0 c_0}{N_T(0)\hbar\omega_j}}A_j(z)$ are the corresponding dimensionless field amplitude with $N_T(0)$ being the total photon flux at $z=0$. Here, the intensity of field at $\omega_j$ is $I_j(z)=2n_r(\omega_j)\epsilon_0 c_0 A_j(z) A_j^*(z)$ so the total intensity of the light in the crystal gives $I_T(z)=\sum\limits_{j}I_j(z)$ and the photon flux is then defined as $N_T(z)=\sum\limits_j N_j(z)=\sum\limits_j I_j(z)/{\hbar \omega_j}$.

In the following, we take $\Delta\beta=0$ for the simplicity to explore the $\mathcal{PT}$-symmetry phase transitions under the phase-matching condition. We define ${g}(z) \equiv \Gamma {u_b}(z)$ and $ \{ {\begin{array}{*{20}{c}}
{{{\tilde u}_a}(z),}&{{{\tilde u}_c}(z)}
\end{array}} \} \equiv e^{\alpha z/2} \{u_a(z),u_c(z)\}$, so Eq.~(\ref{eq2}) gets converted to
\begin{equation}\label{eq3}
 - i\frac{d}{{dz}}\left[ {\begin{array}{*{20}{c}}
{{{\tilde u}_a}(z)}\\
{{u_b}(z)}\\
{{{\tilde u}_c}(z)}
\end{array}} \right] = \left[ {\begin{array}{*{20}{c}}
{ - i\alpha /2}&0&{g (z)}\\
{\Gamma u_c^*(z)}&0&0\\
{{g ^*}(z)}&0&{i\alpha /2}
\end{array}} \right]\left[ {\begin{array}{*{20}{c}}
{{{\tilde u}_a}(z)}\\
{{u_b}(z)}\\
{{{\tilde u}_c}(z)}
\end{array}} \right]\equiv \tilde H_{QPA}(z)\left[ {\begin{array}{*{20}{c}}
{{{\tilde u}_a}(z)}\\
{{u_b}(z)}\\
{{{\tilde u}_c}(z)}
\end{array}} \right].
\end{equation}
$\tilde H_{QPA}(z)$ represents the dynamical Hamiltonian of such nonlinear QPA system with three eigenvalues $\tilde \lambda_{0,\pm}  = 0$ and $ \pm \sqrt {{{\left| {{g }(z)} \right|}^2} - {{\left( {\frac{\alpha }{2}} \right)}^2}} $, together with corresponding eigenvectors $\tilde v_{0,\pm}(z) = {\left[ {\begin{array}{*{20}{c}}
{\begin{array}{*{20}{c}}
{0,}&1,&{0}
\end{array}}
\end{array}} \right]^T}$ and $\frac{1}{{\sqrt 2 }}{\left[ {\begin{array}{*{20}{c}}
{1,}&0&{\frac{{ - i\left| \alpha  \right| \pm \sqrt { - {\alpha ^2} + {{\left| g  \right|}^2}} }}{{g^*}}}
\end{array}} \right]^T}$, respectively. One can find that there exists a pair of EPs at 
\begin{equation}\label{eq4}
\alpha=\pm 2|g(z)|,
\end{equation}
labeled as $z_{_{EP}}$. The $\mathcal{PT}$-symmetry phase of this system can be analyzed based on the basis $\{ \tilde u_a, \tilde u_c \}$, where the amplitude of  $u_b(z)$ (or $g(z)$) determines the $\mathcal{PT}$-symmetry phase transitions while the field propagates along $z$, i.e., oscillations at the $\mathcal{PT}$-symmetry phase and amplifications at the $\mathcal{PT}$-broken phase \cite{el2015ol,flemens2022prl}.

\section{Analysis and results}
\subsection{ Features of $\mathcal{PT}$-symmetry phase transitions under different parametric conditions}

We now show the appearance of multiple $\mathcal{PT}$-symmetry phase transitions under different parametric conditions in the QPA system in Fig.~\ref{fig1}. We first analyze the density ratio $\eta(\alpha,z)  \equiv \tilde n_a/(\tilde n_a + \tilde n_c) = n_a/(n_a + n_c)$ on the basis $\{ \tilde u_a, \tilde u_c \}$, and also the imaginary part of the eigenvalue $\tilde\lambda_\pm$, where $n_i(z)\equiv|u_i(z)|^2$ gives the dimensionless photon flux density for each field. Here, $\eta(\alpha,z)$ labels the flux density ratio between the pump density and idler density that represents signatures of $\mathcal{PT}$-symmetry phase. To plot $\eta(\alpha,z)$ and $|$Im$\tilde\lambda_\pm|$ versus the propagation distance $z$, we numerically solve Eq.~(\ref{eq2}) with different choices of the absorption coefficient $\alpha$, and the initial combination of densities, i.e., $\left\{n_a(0),n_b(0),n_c(0)\right\}$ ($n_a(0)+n_b(0)+n_c(0) = 1$). Hence $|$Im$\tilde\lambda_\pm|$ (dependent on $u_b(z)$) and $\eta(\alpha,z)$ (dependent on $n_a$ and $n_c$) can be obtained.

\begin{figure}
\centering
\includegraphics[width=12cm]{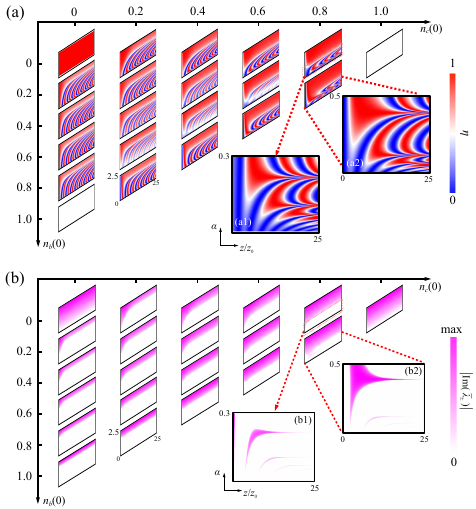}
\caption
{\label{fig2} $\mathcal{PT}$-symmetry phase transitions of QPA system under different initial input conditions $\{ n_a(0),n_b(0),n_c(0) \}$. (a) The evolution of density ratio $\eta(\alpha,z)$. In each subfigure, the horizontal axis is dimensionless propagation distance $z$ and the vertical axis is linear absorption coefficient $\alpha$. (a1) and (a2) Zoom-in plots for $\eta(\alpha,z)$ under $\{ n_a(0),n_b(0),n_c(0) \}=\{ 0.2,0,0.8 \}$ and $\{ n_a(0),n_b(0),n_c(0) \}=\{ 0,0.2,0.8 \}$, respectively. (b) Imaginary eigenvalues $|$Im$\tilde\lambda_\pm|$ where white (purple) regions corresponds to the $\mathcal{PT}$-symmetry(-broken) phase, respectively. (b1) and (b2) Zoom-in plots for $|$Im$\tilde\lambda_\pm|$ under $\{ n_a(0),n_b(0),n_c(0) \}=\{ 0.2,0,0.8 \}$ and $\{ n_a(0),n_b(0),n_c(0) \}=\{ 0,0.2,0.8 \}$, respectively.
}
\end{figure}

Figs.~\ref{fig2}(a)-(b) plot respectively, the variations of density ratios $\eta(\alpha,z)$ and imaginary eigenvalues $|$Im$(\tilde\lambda_\pm)|$ on different $n_c(0)$ (from 0 to 1 horizontally) and $n_b(0)$ (from 0 to 1 vertically), where each subfigure shows the result at different $z$ under various $\alpha$. One can see that Fig.~\ref{fig2} exhibits different characteristics of $\mathcal{PT}$-symmetry phase transitions during the light propagation when photon conversion under PDC ($\omega_a \to \omega_b+\omega_c$) and its inverse conversion ($\omega_b+\omega_c \to \omega_a$) occur simultaneously, i.e., $\eta(\alpha,z)$ oscillates as a function of $z$ at a fixed $\alpha$. Most subfigures in Fig.~\ref{fig2}(a) show such oscillation features for small $\alpha$. For example, subfigures in the second row at $n_b(0)=0.2$ exhibit the features that photon conversions continuously happen as the absorption on photons at $\omega_c$ is small. Subfigures in Fig.~\ref{fig2}(b) also reflect such feature. The aforementioned regions in which $\eta(\alpha,z)$ oscillates in Fig.~\ref{fig2}(a) corresponds to the white regions in Fig.~\ref{fig2}(b) where $|$Im$(\tilde\lambda_\pm)| \to 0$, meaning the system is under the $\mathcal{PT}$-symmetry phase and supports two different $|$Re$(\tilde\lambda_\pm)| \neq 0$ resulting in oscillating dynamics. As the obvious distinction, one notes that for the choice of larger $\alpha$, $\eta(\alpha,z)$ saturates towards unity in Fig.~\ref{fig2}(a), indicating $n_c\to 0$. This result corresponds to regions where $|$Im$(\tilde\lambda_\pm)| \neq 0$ but $|$Re$(\tilde\lambda_\pm)| \to 0$ in Fig.~\ref{fig2}(b) as the system goes into the $\mathcal{PT}$-broken phase, where the population of all three types of photons becomes stable. The critical point of $\alpha$ where the phase transition occurs at large distance $z$ gradually shifts to smaller value with $n_c(0)$ increasing. For example, for the case of $n_b(0)=0.2$, the increase of $n_c(0)$ leads to a larger portion of the input idler photons at $\omega_c$ that promotes the photon inversion, which results in the fast inverse conversion to pump photons at $\omega_a$ under the absorption of idler photons. Hence pump photons become dominant and the region of $\eta(\alpha,z)\to 1$ occupies larger region in the plots in Fig.~\ref{fig2}(a), which corresponds to the larger region of non-zero $|$Im$(\tilde\lambda_\pm)|$ in Fig.~\ref{fig2}(b).

As the most striking feature, for cases at  $n_c(0)=0.8$ with either $n_a(0)=0.2$, $n_b(0)=0$ or $n_a(0)=0$, $n_b(0)=0.2$ (zoomed-in in Figs.~\ref{fig2}(a)-(b), respectively), there are multiple $\mathcal{PT}$-symmetry phase transitions in the small $\alpha$ regime originated from the competition between PDC and its inverse conversion. As the immediate consequence at large $z$, with the continual absorption and the inverse conversion, this competition makes signal photons decrease and so Figs.~\ref{fig2}(b1)-(b2) depict sharp peaks, which is also reflected as multiple staggered stripes in Figs.~\ref{fig2}(a1)-(a2). Other details can be seen in Supplementary Material.

\subsection{Relationships between $\mathcal{PT}$-symmetry phase transitions and photon flux densities}

\begin{figure}[H]
\center
\includegraphics[width=12cm]{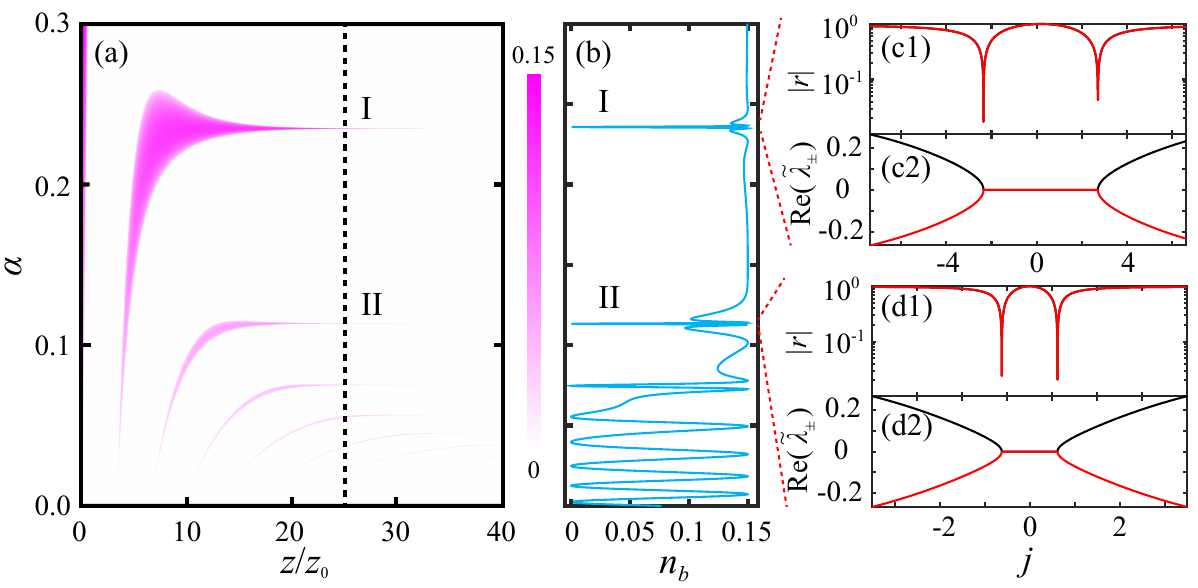}
\caption{\label{fig3} (a) Imaginary eigenvalues $|$Im$\tilde\lambda_\pm|$ under the initial condition $n_a(0) =0.15 $ and $n_c(0) =0.85$. The dashed line corresponds to $z=25z_0$, with I and II in the vicinities of two transition points. (b) Photon flux density $n_b$ as a function of $\alpha$ (the vertical axis) plotted at $z=25z_0$. Two sharp dips of $n_b$ can be seen at similar regimes I and II of $\alpha$ respectively. (c1) and (d1) Phase rigidity $|r|$, (c2) and (d2) Re$(\tilde\lambda_{\pm})$ of two dips versus the zoomed-in $\alpha$ where the horizontal axis corresponds to $j=\left[\alpha-\alpha_{\rm{I_0(II_0)}}\right] / \Delta\alpha$ with $\alpha_{\rm{I_0(II_0)}}=0.235718(0.113317)$ and $\Delta\alpha=10^{-6}$. In (c2) and (d2) black (red) lines denote to the eigenvalues Re$(\tilde\lambda_\pm)$.
}
\end{figure}

To investigate the relationship between $\mathcal{PT}$-symmetry phase transitions and photon flux densities, we consider in detail the case of $\{n_a(0),n_b(0),n_c(0)\}=\{0.15,0,0.85\}$ and plot the imaginary eigenvalues $|$Im$(\tilde\lambda_\pm)|$ in Fig.~\ref{fig3}(a). One can see similar multiple phase transitions as those in Figs.~\ref{fig2}(b1)-(b2). At the distance $z=25z_0$, we plot the photon flux density of the signal density $n_b(z=25z_0)$ in Fig.~\ref{fig3}(b). One can see depletion dips of $n_b$ at each value of $\alpha$ that the phase transitions occur at $\alpha_{\rm{I_0}}$=0.235718 and $\alpha_{\rm{II_0}}$=0.113317 (labeled as the dips I and II) respectively in Fig.~\ref{fig3}(a). Such dips are more sharper for larger $\alpha$, i.e., $\alpha>0.1$, while for $\alpha<0.1$, the photon flux of $n_b$ exhibits oscillation feature. Note that the number of significant digits in $\alpha$ may affect the precision in the non-Hermitian sensing proposed in the next section, so we leave the number of the significant digits in $\alpha$ upto six here to see its sensing capability in the theoretical limit. Decreasing the number of the significant digits in $\alpha$ will not affect the sensing mechanism with EPs but might decrease the sensing precision. 


Other than the imaginary eigenvalue plotted in Fig.~\ref{fig2}(b), we also calculate the phase rigidity $|r|=1/\left\langle {\tilde v_\pm|\tilde v_\pm} \right\rangle $ \cite{ding2018prl,bai2022nsr,bai2023prl} and include the real eigenvalue Re$(\tilde\lambda)$, both of which are plotted in Figs.~\ref{fig3}(c1)-(c2) and (d1)-(d2) with the zoom-in $\alpha$ in the vicinities of two phase transition dips (I \& II). We can see the phase rigidity vanishes at the EPs, which can also been seen in the coalescence of the real eigenvalues in Figs.~\ref{fig3}(c2)-(d2). It is worthy to note that the difference $\Delta\alpha_{EP}$ between two transition points in Figs.~\ref{fig3}(c1)-(d2) is not equivalent to the full width at half maximum of dips of $n_b$ in Fig.~\ref{fig3}(b) owing to non-immediate reaction to convert the signal density in such small scopes of $\mathcal{PT}$-broken phase. Nevertheless, the sharp dip of $n_b$ between phase transitions holds the potential capability for the sensing under the small variation of surrounding conditions. In other words, such multiple phase transitions can provide a new perspective to explore non-Hermitian sensing in a nonlinear system.

\subsection{Feasibility of non-Hermitian sensing from multiple phase transitions}


As an example, we consider the problem of sensing the nearly homogeneousness of the doped absorber along $z$ direction in the nonlinear system. Small spatial perturbation on $\alpha$ may be not easy to be captured from a simple absorption measurement. In this case, the sharp dips in the photon flux density $n_b$ provide a key candidate for sensing quantitatively. The idea is to tune the system operated at the regime between phase transitions (i.e., the dip I or II in Fig.~\ref{fig3}(a)) and then measure $n_b$. For an absolute homogeneous distribution of $\alpha$ in the $z$ direction, one expects a measurement of the variation of $n_b$ to be zero. Nevertheless, due to the extreme sensitivity from $\mathcal{PT}$-symmetry phase transitions, small perturbation in $\alpha$ can result in different measurements of $n_b$.

\begin{figure}[htb]
\center
\includegraphics[width=12cm]{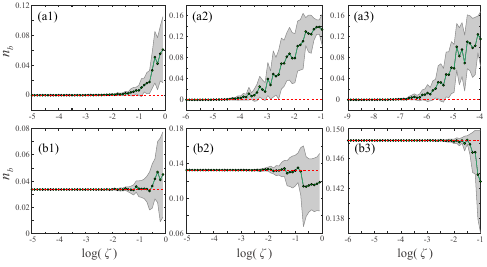}
\caption{\label{fig4}
Photon flux densities of the signal photon ($n_b$) calculated with different amplitudes of perturbation ($\zeta$) with the average $\alpha$ chosen as (a1)-(a3) $\alpha_{\rm{I_0}}$ for the dip I and (b1)-(b3) $\alpha_{\rm{II_0}}$ for the dip II in Fig.~\ref{fig3}(a). Different distances of the nonlinear crystal ($z_{max}$) are taken in simulations, i.e., (a1) and (b1) $z_{max}=10z_0$, (a2) and (b2) $z_{max}=20z_0$, (a3) and (b3) $z_{max}=30z_0$. The grey shadow denotes the standard deviations from twenty sets of simulation results under the same parameters but different randomness. The red dashed line is the result of $n_b$ in the homogeneous crystal ($\zeta=0$). The parameters are set to the condition of exact phase-matching at the center of the dip I ($\alpha_{\rm{I_0}}$).
}
\end{figure}


To calibrate the perturbation in $\alpha$, we define the amplitude of perturbation $\zeta$ and randomly set $\alpha\in[(1-\zeta)\alpha_0, (1+\zeta)\alpha_0]$ continuously in the nonlinear system, where $\alpha_0$ is the value of the homogeneous absorption (see more details in Supplementary Material). Figs.~\ref{fig4}(a1)-(a3) and (b1)-(b3) show the statistics of the output $n_b$ for the choice of different $\zeta$ with the choice of $\alpha_{\rm{I_0(II_0)}}$, i.e., the dips I and II in Fig.~\ref{fig3}(b). Three different distances of the nonlinear crystal ($z=10z_0,20z_0,30z_0$) are taken corresponding to the same transition regime in Fig.~\ref{fig3}(a), where larger distance gives sharper transition on $\alpha$ but less sensitivity on $|$Im$(\tilde\lambda_\pm)|$. For each choice of parameters, we count twenty sets of simulation data with the random set of $\alpha$. In Figs.~\ref{fig4}(a1)-(a3), one can see that for small perturbation $\zeta$, three choices of the crystal distance all give a stable measurement of $n_b$. The threshold for $\zeta$ (labeled as $\zeta_0$) that the measurement of $n_b$ becomes unstable occurs at different values for three cases. Namely,  the threshold $\zeta_0 \approx 10^{-1.5}, 10^{-4}, 10^{-6.5}$ for $z=10z_0$, $20z_0$, $30z_0$, respectively. It is clear to see that, in this choice of $\alpha_0$, the longer distance the crystal is, the better sensitivity for the spacial perturbations on $\zeta$ we obtain. For $z=30z_0$ one can expect to measure the variation of the absorber upto $10^{-7}$ in Fig.~\ref{fig4}(a3). One shall note that here we adjust parameters to satisfy the exact phase-matching condition at the center of the dip I and hence for other $\alpha$ there are small phase-mismatching on wave vectors under the spacial perturbations. Other choices of parameters for the phase-matching condition at other values of $\alpha$ can be found in Supplementary Material. As a comparison, under this phase-matching condition for the dip I, we also study cases with the choice of $\alpha_{\rm{II_0}}$ at the dip II. In Figs.~\ref{fig4}(b1)-(b3), the threshold for $\zeta_0$ occurs at larger perturbation for all three crystal distances, meaning the less sensitivity for this case.

\begin{figure}[htb]
\center
\includegraphics[width=12cm]{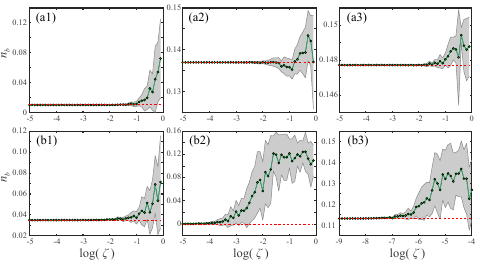}
\caption{\label{fig5} 
Photon flux densities of the signal photon ($n_b$) calculated with different amplitudes of perturbation ($\zeta$) with the average $\alpha$ chosen as (a1)-(a3) $\alpha_{\rm{I_0}}$ for the dip I and (b1)-(b3) $\alpha_{\rm{II_0}}$ for the dip II in Fig.~\ref{fig3}(a). Different distances of the nonlinear crystal ($z_{max}$) are taken in simulations, i.e., (a1) and (b1) $z_{max}=10z_0$, (a2) and (b2) $z_{max}=20z_0$, (a3) and (b3) $z_{max}=30z_0$. The grey shadow denotes the standard deviations from twenty sets of simulation results under the same parameters but different randomness. The red dashed line is the result of $n_b$ in the homogeneous crystal ($\zeta=0$), The parameters are set to the condition of exact phase-matching at the center of the dip II ($\alpha_{\rm{II_0}}$).
}
\end{figure}

One can also switch the phase-matching condition, i.e., to set it at the center of the dip II, and perform the simulations for different perturbations on $\zeta$. The corresponding results are plotted in Fig.~\ref{fig5}. In this case, for the choice of $\alpha_{\rm{II_0}}$ at the dip II, we can see $\zeta_0=10^{-1.5},10^{-4},10^{-7}$ for $z=10z_0,20z_0,30z_0$, respectively, while for the choice at the dip I with the phase-mismatching, $\zeta_0$ is roughly near $10^{-1.5}$ for all three crystal distances.


Although one can tell the threshold value of $\zeta_0$ when it varies away from its stable value in Figs.~\ref{fig4}-\ref{fig5}, it is useful to give a quantitative measurement. Thus, we define the contrast ratio $\Lambda = 1 - \frac{{\bar n_b^{\zeta  = {\zeta _T}}({z_{max }})}}{{{n_a}(z = 0) - n_b^{\zeta  = 0}({z_{max }})}}$, where $n_a(z=0)$ is the input photon flux density for the pump and $n_b^{\zeta=0}(z_{max})$ give the output signal photon flux density in the case of no perturbation on the absorption coefficient, and also $\zeta_T$, i.e., the perceived precision, $\bar n_b^{\zeta  = {\zeta _T}}({z_{max }}) = n_b^{\zeta  = 0}({z_{max }}) + \left[ {{n_a}(z = 0) - n_b^{\zeta  = 0}({z_{max }})} \right] \cdot 5\%$. If the contrast ratio satisfies $\Lambda \to 1$, the system holds a sharp dip for the relation between $n_b$ and $\alpha$ and the sensing is supported to be distinguishable.

\begin{figure}[htb]
\center
\includegraphics[width=11cm]{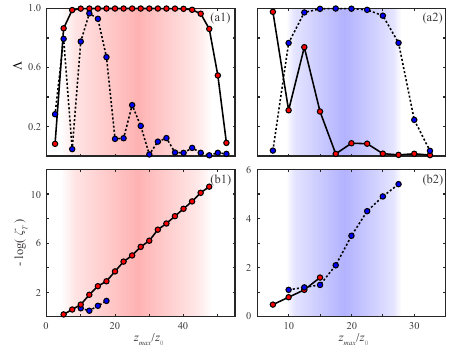}
\caption{\label{fig6} The contrast ratio $\Lambda$ and the transitive perturbation rate $\zeta_T$ versus different nonlinear crystal distances ($z_{max}$) from taking parameters meeting phase-matching conditions (a1), (b1) at the centers of the dip I and (a2), (b2) at the center of the dip II. Red and blue dots denote the results using $\alpha_{\rm{I_0}}$ at the dip I and $\alpha_{\rm{II_0}}$ at dip II respectively. The red shadows in (a1) and (b1) (the blue shadows in (a1) and (b2)) refers to the effective perceiving scopes for the dips I and II under perturbations, respectively. 
}
\end{figure}

 
We plot $\Lambda$ and $\zeta_T$ with the propagation distance $z_{max}$ in Fig.~\ref{fig6}. In Figs.~\ref{fig6}(a1)-(b1) for the phase-matching condition at the dip I, one can see that the contrast ratio for the dip I is close to unity from $z_{max}=5z_0$ to $47.5z_0$ because of a sharp peak in the $\mathcal{PT}$-symmetry phase transition. In this regime, the system is suitable for sensing. The corresponding quantity $-$log$(\zeta_T)$ increases linearly and we can obtain more sensitive precision at a larger distance. In particular, for $z_{max}=47.5z_0$, we have $\Lambda\sim 0.6$ and $\zeta_T \sim 10^{-11}$, indicating ultra-sensitive measurement for the small perturbation on the absorption coefficient. As the comparison, when we choose $\alpha$ near the dip II, $\Lambda$ drops quickly after $z_{max}>17.5z_0$, with $\zeta_T$ upto $\sim 10^{-1}$ in Fig.~\ref{fig6}(b1). Similar results can be found in Figs.~\ref{fig6}(a2)-(b2), where we choose the phase-matching at the dip II. In this case, we obtain a larger contrast when $\alpha$ is chosen at the dip II, with $\zeta_T$ reaching $10^{-5}$ at $z_{max}=27.5z_0$. One can see that the sensing can be more sensitive when we tune the phase-matching at the dip I because of the larger transition regime shown in Fig.~\ref{fig3}(a). More discussions on sensing results of cases when one sets the phase-mismatching at centers of dips are given in Supplementary Material.

\section{Conclusion}
In summary, we have shown that the dynamical evolution of the QPA is controlled through absorption parameter of the system and its spatial distribution. The complexity of nonlinear optical interactions and internal competition for photon conversions supports the abundance of $\mathcal{PT}$-symmetry phase transitions. We reveal that a small variation of surrounding conditions can directly cause evident changes in output photon flux densities at the phase transition around EPs, which enlightens towards sensing for the homogeneousness of the absorber. We showcase the ultra-sensitive measurement of small perturbations on the distribution of absorbers utilizing the EP and transition points. The proposed novel scheme therefore holds important promise for investigating multiple $\mathcal{PT}$-symmetry phase transitions as well as demonstrating EP sensing in nonlinear optics, and these sensitive dips can be potentially extended to evaluate thermal fluctuation due to rapidly changing light field distributions \cite{gavartin2012NN,monzel2016JPAP,lu2019APR}.

\section{Acknowledgments}
This research was supported by the National Natural Science Foundation of China (12122407 and 12192252) and the National Key Research and Development Program of China (No. 2023YFA1407200 and No. 2021YFA1400900). V.Y. acknowledges support of the NIH (R21CA269099, R21GM142107, and R01GM127696) and AFOSR (FA9550-20-1-0366). L.Y. thanks the sponsorship from Yangyang Development Fund.

\bibliography{Ref}

\end{document}